%% file: article.tex
\documentclass[
10pt, % Main document font size
a4paper, % Paper type, use 'letterpaper' for US Letter paper
oneside, % One page layout (no page indentation)
headinclude,footinclude, % Extra spacing for the header and footer
BCOR5mm, % Binding correction
]{scrartcl}

\input{structure.tex} % Include the structure.tex file which specified the document structure and layout

\title{\normalfont\spacedallcaps{A cascade network for Detecting COVID-19 using chest x-rays}} % The article title

\author{\small\spacedlowsmallcaps{Dailin Lv \textsuperscript{1}, Wuteng Qi\textsuperscript{1}, Yunxiang Li\textsuperscript{1}, Lingling Sun\textsuperscript{1}, Yaqi Wang\textsuperscript{1}}} % The article author(s) - author affiliations need to be specified in the AUTHOR AFFILIATIONS block

\date{} % An optional date to appear under the author(s)

\begin{document}

%----------------------------------------------------------------------------------------
%	HEADERS
%----------------------------------------------------------------------------------------

\renewcommand{\sectionmark}[1]{\markright{\spacedlowsmallcaps{#1}}} % The header for all pages (oneside) or for even pages (twoside)
\lehead{\mbox{\llap{\small\thepage\kern1em\color{halfgray} \vline}\color{halfgray}\hspace{0.5em}\rightmark\hfil}} % The header style

\pagestyle{scrheadings} % Enable the headers specified in this block

%----------------------------------------------------------------------------------------
%	TABLE OF CONTENTS & LISTS OF FIGURES AND TABLES
%----------------------------------------------------------------------------------------

\maketitle % Print the title/author/date block

\setcounter{tocdepth}{2} % Set the depth of the table of contents to show sections and subsections only

%\tableofcontents % Print the table of contents

%\listoffigures % Print the list of figures

%\listoftables % Print the list of tables

%----------------------------------------------------------------------------------------
%	ABSTRACT
%----------------------------------------------------------------------------------------

\section*{Abstract} % This section will not appear in the table of contents due to the star (\section*)

With the spread of pneumonia caused by SARS-CoV-2 around the world, as of April 22, more than 2.5 million people had been diagnosed with pneumonia in the world with a mortality rate of nearly 7\%. This poses an unprecedented challenge to medical resources and prevention and control measures around the world. Covid-19 attacks not only the lungs, making it difficult to breathe and life-threatening, but also the heart, kidneys, brain and other vital organs of the body, with possible sequela. At present, the detection of COVID-19 needs to be realized by the reverse transcription-polymerase Chain Reaction (RT-PCR). However, many countries are in the outbreak period of the epidemic, and the medical resources are very limited. They cannot provide sufficient numbers of gene sequence detection, and many patients may not be isolated and treated in time. To assist doctors to diagnose and increase efforts to inspect, a rapid and effective detection method is particularly important. Given this situation, we researched the analytical and diagnostic capabilities of deep learning on chest radiographs and proposed Cascade-SEMEnet which is cascaded with SEME-ResNet50 and SEME-DenseNet169. The two cascade networks of Cascade - SEMEnet both adopt large input sizes and SE-Structure and use MoEx and histogram equalization to enhance the data. We first used SEME-ResNet50 to screen chest X-ray and diagnosed three classes: normal, bacterial, and viral pneumonia. Then we used SEME-DenseNet169 for fine-grained classification of viral pneumonia and determined if it is caused by COVID-19. To exclude the influence of non-pathological features on the network, we preprocessed the data with U-Net during the training of SEME-DenseNet169. The results showed that our network achieved an accuracy of 85.6\% in determining the type of pneumonia infection and 97.1\% in the fine-grained classification of COVID-19. We used Grad-CAM to visualize the judgment based on the model and help doctors understand the chest radiograph while verifying the effectivene.

%----------------------------------------------------------------------------------------
%	AUTHOR AFFILIATIONS
%----------------------------------------------------------------------------------------

\let\thefootnote\relax\footnotetext{\textsuperscript{1} \textit{CAD institute,Hangzhou Dianzi University,Hangzhoou,China}}

%----------------------------------------------------------------------------------------

\newpage % Start the article content on the second page, remove this if you have a longer abstract that goes onto the second page

%----------------------------------------------------------------------------------------
%	INTRODUCTION
%----------------------------------------------------------------------------------------

\section{Introduction}

SARS-CoV-2 that spread all over the world and poses a deadly threat to people's health has now caused a global pandemic. Epidemiology unit of COVID-19 emergency response mechanism, Chinese center for disease control and prevention issued a paper, indicating that SARS-CoV-2 is more infectious than SARS and MERS\cite{1}. As of April 22, 2020, more than 2.5 million people had been diagnosed with pneumonia in the world and nearly 180,000 people died of it. The death rate calculated by these data is as high as 7\%. Recent studies\cite{4} have found that COVID-19 not only has devastating effects on human lung tissues, but also attacks vital organs such as heart, blood vessels, kidneys, gastrointestinal tract, eyes, and brain, causing very serious consequences. Survivors of severe COVID-19 patients may also be at risk of disability\cite{5}.

Detection of COVID-19 in many countries is confirmed by gene sequencing of breath or blood samples that used RT-PCR. However, the epidemic is in an outbreak period and many countries are not able to provide sufficient numbers of gene sequencing, which may mean that many patients cannot be quickly identified and receive proper treatment. The team led by Zhong Nanshan who is an academician of Chinese Academy of Engineering\cite{7} investigated the data of 1099 lab-confirmed COVID-19 patients from 552 hospitals in 30 provinces, autonomous regions and municipalities in mainland China as of January 29, 2020, and found that approximately 86\% of the patients had abnormal results in chest imaging after analysis. After analyzing the chest images of 81 COVID-19 patients in Wuhan, Heshui Shi\cite{2} et al. Found that COVID-19 pneumonia manifests with chest CT imaging abnormalities, even in asymptomatic patients, with rapid evolution from focal unilateral to diffuse bilateral ground-glass opacities that progressed to or co-existed with consolidations within 1-3 weeks. Therefore, combining the assessment of imaging features with the results of clinical and laboratory examinations can facilitate the early diagnosis of COVID-19 pneumonia\cite{2}. In the clinical diagnosis of pneumonia, chest imaging has a good effect. This is not the first time that the diagnosis of pneumonia is correlated with chest imaging features in medicine. In previous studies on pneumonia imaging, doctors were able to diagnose whether pediatric pneumonia is bacterial pneumonia by chest X-ray\cite{3}.

In a large number of previous experiments, we found that deep learning performed very well in chest imaging, and deep convolutional neural network could accurately diagnose whether a patient had pneumothorax\cite{8}.U-net\cite{9} based on Resnet50 can also accurately identify the range of double lungs in the chest radiograph and infer the pneumothorax area\cite{10}. It can be seen that the convolutional neural network (CNN) in deep learning can effectively extract the characteristics of lung lesions after accurate data labeling, and we can use the characteristics in the subsequent network structure to achieve the purpose of diagnosing the lesions. Daniel Shu Wei Ting et al wrote in Nature\cite{6}that digital technologies including deep learning can be used to remedy the COVID-19 epidemic. Therefore, Given the recent sudden outbreak of COVID-19, we conducted a study on automatic diagnosis of pneumonia by deep learning.

When respiratory symptoms occur, using lung imaging to judge the type of lung infection is particularly important, which directly affects the next treatment. If doctors can determine the type of lung infection by chest X-ray, then features in different lesions of the chest X-ray have a good chance to be extracted by the convolution neural network (CNN) layer. The neural network model can according to these characteristics, to assist the doctor diagnosed the type of pneumonia. Our main research is divided into two parts: the first part is to train the neural network model to perform a kind of pre-diagnosis on the chest image to distinguish whether there is pathological change on lungs and whether the lung lesions are caused by the bacterial or viral infection. The second part is about viral pneumonia, using AI to classify the images of viral pneumonia in a fine-grained manner.   Pneumonia caused by virus infection often has very serious consequences. MERS, SARS, and now COVID-19 in the global scope all have extremely high mortality and infectivity. Moreover, different types of pneumonia also have their own characteristics.

\begin{figure}[!t]
	\centerline{\includegraphics[width=\columnwidth]{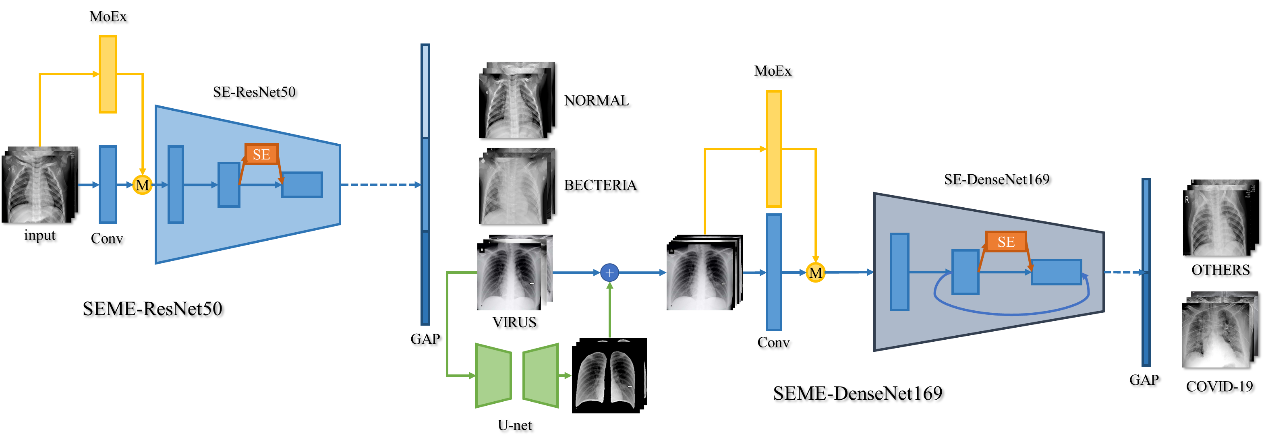}}
	\caption{The overall architecture of Cascade-SEMEnet pneumonia detection network is composed of SEME-ResNet50 and SEME-DenseNet169. U-Net is used to remove non pathological features in X-ray films.}
	\label{fig1}
\end{figure}

Our training on the network is based on open data sets \cite{18} and \cite{19}, which are X-ray data sets conclude bacterial pneumonia, viral pneumonia and normal lung, and COVID-19 data sets about bacterial pneumonia. This paper discusses the diagnostic effect of VGG19, ResNet50, DenseNet169 , and their optimized structures on chest radiographs. We propose Cascade-SEMEnet (Figure\ref{fig1}) and evaluate the diagnostic effect of our network on chest radiographs for pneumonia. We modified the pooling layer according to the receptive field of the network and added the attention mechanism, which can effectively improve the judgment effect of the network.

Due to the small amount of data, in order to eliminate the interference of data as much as possible, we used U-Net\cite{9} to segment the images of viral pneumonia in advance, and only the double-lung images were retained during the training. We used affine transformation, histogram enhancement and other data augmentation methods to visually amplify the data, so as to increase the diversity of chest image distribution. Meanwhile, we also adopted MoEx\cite{22}, which has an orthogonal effect with traditional data augmentation methods. Different from the traditional method of data enhancement, MoEx integrates the characteristics of two different categories of data and interpolates them in the classification label. This method makes data augmentation in the image feature space, and its improvement can be superimposed on the existing enhancement effect.

In the evaluation, the confusion matrix of the network was analyzed and calculated, and the ROC curves of each network were compared to verify the performance of the network. We used Grad-CAM method to calculate the weighted sum of the feature map in each convolution layer of the input network image, and obtained a thermal map to interpret the classification results, so as to verify the reliability of the trained network.

in this paper, COVID-19 and other kinds of pneumonia in chest radiography are studied in the second part, and the application of deep learning in lung imaging is investigated. The third part lists the methods we used in the research process, including model building method, data enhancement method and the visual method of the discrimination basis. We carried out the experiment in the fourth part, and analyzed the experimental results in the fifth part, and finally came to the conclusion.

%----------------------------------------------------------------------------------------
%	Relate work
%----------------------------------------------------------------------------------------
\section{Relate work}
\begin{figure}[!t]
	\centerline{\includegraphics[width=\columnwidth]{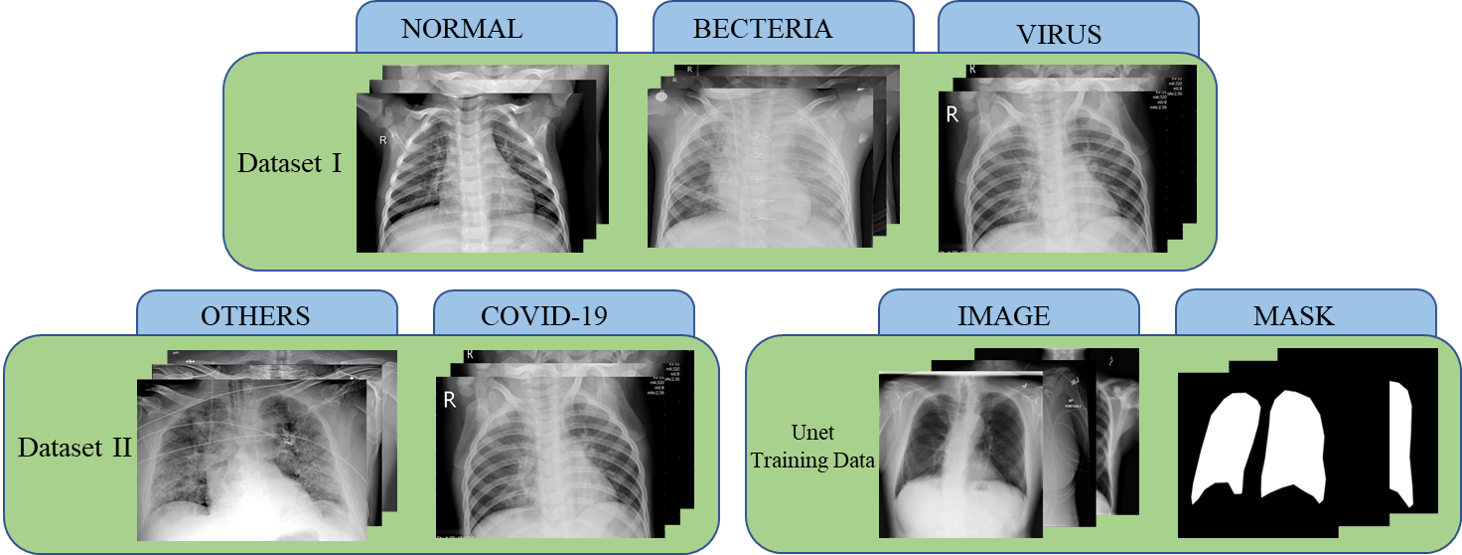}}
	\caption{Display of dataset1 and dataset2.}
	\label{fig2}
\end{figure}

All data in this article are publicly available online. The data consists of two parts: we obtained the data set of the first part from\cite{18}, which contains three categories that 5859 chest X-rays were taken from patients with normal lungs, bacterial pneumonia, and viral pneumonia. The second part of the data is obtained from\cite{19}. This data set collects and arranges the chest images of patients with pneumonia, such as MERS, SARS, COVID-19, etc. from various publications. It is still being updated. Since there is less data on other types of viral pneumonia, we only used chest radiographs from COVID-19 patients. In the experiment, we used U-Net to cut the lung area in the chest radiographs, and the data of training U-Net were the mask of the lung area in the chest radiographs and the chest radiographs. We obtained this part of the data from\cite{20}. Figure\ref{fig2} is the data display, and Table\ref{tab1} is the statistics and division of these data.

% Please add the following required packages to your document preamble:
\begin{table}
	\label{tab1}
	\centering
	\caption{Statistics and division of Dataset1, Dataset2 and training data of U-Net.}
	\small
	\renewcommand\tabcolsep{2.0pt} % 调整表格列间的宽度
	\begin{tabular}{cccccccccc}\toprule[1pt] 
	{\color[HTML]{000000} } & \multicolumn{3}{c}{{\color[HTML]{000000} \textbf{Dataset I}}} & {\color[HTML]{000000} } & \multicolumn{2}{c}{{\color[HTML]{000000} \textbf{Dataset II}}} & {\color[HTML]{000000} } & \multicolumn{2}{c}{{\color[HTML]{000000} \textbf{U-Net Training   Data}}} \\ \cline{2-4} \cline{6-7} \cline{9-10}
	{\color[HTML]{000000} } & {\color[HTML]{000000} \textbf{normal}} & {\color[HTML]{000000} \textbf{becteria}} & {\color[HTML]{000000} \textbf{virus}} & {\color[HTML]{000000} } & {\color[HTML]{000000} \textbf{COVID-19}} & {\color[HTML]{000000} \textbf{others virus}} & {\color[HTML]{000000} } & {\color[HTML]{000000} \textbf{image}} & {\color[HTML]{000000} \textbf{mask}} \\
	{\color[HTML]{000000} \textbf{training}} & {\color[HTML]{000000} 1341} & {\color[HTML]{000000} 2530} & {\color[HTML]{000000} 1345} & {\color[HTML]{000000} } & {\color[HTML]{000000} 105} & {\color[HTML]{000000} 165} & {\color[HTML]{000000} } & {\color[HTML]{000000} 900} & {\color[HTML]{000000} 900} \\
	{\color[HTML]{000000} \textbf{validation}} & {\color[HTML]{000000} 125} & {\color[HTML]{000000} 121} & {\color[HTML]{000000} 74} & {\color[HTML]{000000} } & {\color[HTML]{000000} 10} & {\color[HTML]{000000} 76} & {\color[HTML]{000000} } & {\color[HTML]{000000} 50} & {\color[HTML]{000000} 50} \\
	{\color[HTML]{000000} \textbf{test}} & {\color[HTML]{000000} 125} & {\color[HTML]{000000} 121} & {\color[HTML]{000000} 74} & {\color[HTML]{000000} } & {\color[HTML]{000000} 10} & {\color[HTML]{000000} 75} & {\color[HTML]{000000} } & {\color[HTML]{000000} 50} & {\color[HTML]{000000} 50} \\
	{\color[HTML]{000000} \textbf{overall}} & {\color[HTML]{000000} 1591(27\%)} & {\color[HTML]{000000} 2772(47\%)} & {\color[HTML]{000000} 1493(26\%)} & {\color[HTML]{000000} } & {\color[HTML]{000000} 125(29\%)} & {\color[HTML]{000000} 316(71\%)} & {\color[HTML]{000000} } & {\color[HTML]{000000} 1000} & {\color[HTML]{000000} 1000} \\ \bottomrule[1pt]
\end{tabular}

\end{table}
In the wake of the outbreak, tens of thousands of patients died from pneumonia caused by SARS-CoV-2. Once SARS-CoV-2 invades the body, it seeks cell membrane protein angiotensin-ACE2 (ACE2) as its landing site \cite{21}, and the lung cells that are the main battlefield of SARS-CoV-2 have abundant ACE2. At the same time, SARS-CoV-2 will attack almost all organs throughout the body \cite{4}, such as brains, hearts, kidneys, and other vital organs. Therefore, after a severe patient overcomes a COVID-19, it will also face problems related to recovery \cite{5}. Medical workers have analyzed a considerable amount of the chest images of COVID-19 patients \cite{11}\cite{12}\cite{13}\cite{14}. The lungs of most patients with COVID-19 will show consolidation and Ground-glass opacification. Different from many other types of viral pneumonia, the chest CT of COVID-19 patients will show multiple tiny pulmonary nodules \cite{12}

In April this year, Tej Bahadur Chandra et al published in TMI\cite{15} an analysis of small image patches in the lung area in chest radiographs, demonstrating the promising performance of FCM, KM and other techniques in dividing suspected abnormal areas. Phat Nguyen Kieu et al proposed a deep learning model to detect abnormal sentisy in chest X-ray images in\cite{16} and Fusion rules, for synthesizing the results of the components of the model. Similar to our study, Harsh Sharma et al. \cite{17} proposed a different deep convolutional neural network (CNN) architecture, which extracts features from chest X-ray images and classifies the images to detect whether a person has pneumonia. In a large number of previous experiments, we found that deep learning performed very well in chest imaging, and deep convolutional neural networks could accurately diagnose whether a patient had pneumothorax\cite{8}. Based on the above, this paper studied the automatic diagnosis of COVID-19 by convolutional neural networks. We propose a cascade neural network architecture consisting of SEME-ResNet50 and SEME-DenseNet169. SEME-ResNet50 was used to diagnose the presence and type of infection in the patients’ lungs, while SEME-DenseNet169 was used to classify chest radiographs in a fine-grained manner to determine whether viral pneumonia was caused by SARS-CoV-2. The innovation of this paper lies in:

\begin{enumerate}
\item We propose a Cascade-SEMEnet consisting of a SEME-ResNet50 for detecting the type of lung infection and a DenseNet169 for the subdivision of viral pneumonia, used to diagnose lung disease and the recent outbreak of COVID-19 
\item Use GAP to improve the network structure of ResNet and DenseNet, effectively use the lesion details of the image to increase the receptive field, and add SE-Structure to the network structure, use the attention mechanism for the feature channel. In the network structure based on ResNet50, the accuracy rate of the lung infection type has been increased by 9%.
\item Introduced Contrast Limited Adaptive Histogram Equalization (CLAHE)  that can enhance the random limited contrast of chest radiographs, and a MoEx structure that uses the normalization and enhancement of image features in neural networks, and proved that such image enhancement methods play an active role in classification tasks.
\item Use U-Net to remove the non-pathological features on the chest radiograph avoids learning wrong information in the training process of neural networks and enhances the effectiveness of neural network diagnosis.
\item The Grad-CAM method is used to invert the thermal map of the network on the original image, and the classification basis of the neural network is visualized, which can help doctors better understand the chest radiograph.
\end{enumerate}
%----------------------------------------------------------------------------------------
%	METHODS
%----------------------------------------------------------------------------------------

\section{Methods}

In this article, we studied the VGG19, RseNet50, DenseNet169, and their improvement structure on pneumonia diagnosis task performance, and constructed the Cascade-SEMEnet to make diagnosis and subdivision of pneumonia cases. The structure of Cascade-SEMEnet is shown in Figure\ref{fig1}. It consists of two networks, SEME-ResNet50 and SEME-DenseNet169. SEME-ResNet50 will first make the initial diagnosis of chest radiograph, and the diagnosis results are divided into three categories: namely normal, bacterial pneumonia, and viral pneumonia. Next, SEME-DenseNet169 will conduct a fine-grained classification of viral pneumonia in the initial diagnosis, and find out the chest radiographs of COVID-19 cases. I will describe in detail the approach used to build Cascade-SEMEnet. 

\subsection{Model building}
\begin{figure}[!t]
	\centerline{\includegraphics[width=\columnwidth]{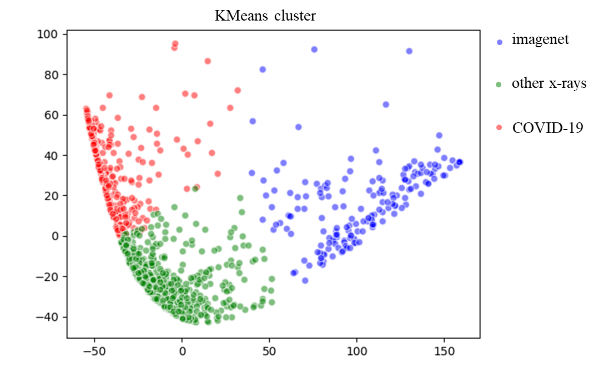}}
	\caption{Kmeans clustering graph of Imagenet, COVID-19 and other X-rays.}
	\label{fig3}
\end{figure}

We selected three classical neural network models, VGG19, ResNet50, and Densenet169, as the basic network models. VGG network structure is simple, stable, and easy to analyze. ResNet, DenseNet and other networks use residual structure and have higher depth and stronger ability to extract features. However, these neural network models are optimized and evaluated on the ImageNet data set. Although they are good at judging natural images of small size, they may not perform as well on the chest image. We extracted randomly 1141 images (its statistics are shown in Table\ref{tab2}) from ImageNet, COVID - 19, and other-xrays. After using hash trac to perform feature dimensionality reduction and K-Means clustering on these images, we found that the feature distribution of ImageNet (Figure\ref{fig3} blue point) is very different from chest radiographs such as COVID-19(Figure\ref{fig3} red and green point). The size and data distribution of the chest radiograph data set are obviously different from the ImageNet data set. Therefore, it is necessary to optimize these networks in the judgment of chest image data.

\begin{table}
	\label{tab2}
	\centering
	\caption{Statistics on the use of K-means clustering.}
	\renewcommand\tabcolsep{5.0pt} % 调整表格列间的宽度
\begin{tabular}{llll}\toprule[1pt]
	\textbf{Dataset} & \textbf{Imagenet} & \textbf{COVID-19} & \textbf{Other X-rays} \\ \hline 
	\textbf{Images} & 500 & 131 & 510 \\ \bottomrule[1pt]
\end{tabular}
	
\end{table}
%\begin{itemize}
%\item a rotation with a random angle between 0 and 90 degrees;
%\item a scale with a random value between 1.1 and 1.3;
%\item addition of gaussian noise to the original image.
%\end{itemize}

%|p{2.25cm}|p{2cm}|l|p{4cm}|p{3cm}|p{2cm}|
\subsubsection{Input image size and receptive field}
\begin{figure}[!t]
	\centerline{\includegraphics[width=\columnwidth]{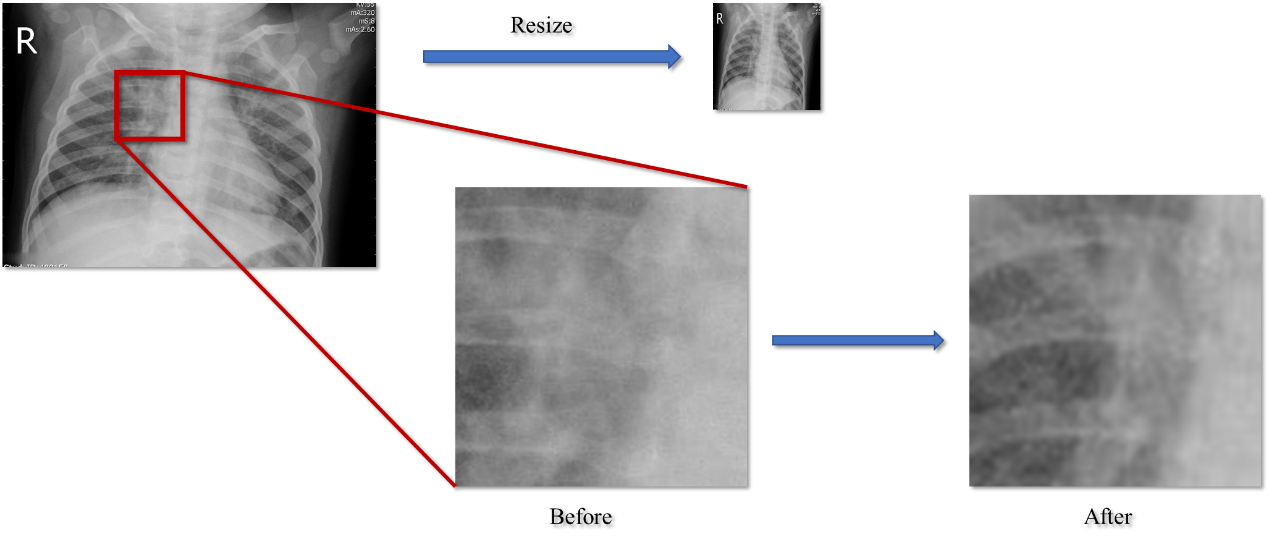}}
	\caption{Resize operation leads to loss of lesion details in the image.}
	\label{fig4}
\end{figure}

Input image size and receptive field
The most intuitive difference between the Imagenet data set and the chest radiograph is the size difference. The image size of the former is only a few hundred times several hundred pixels, while the chest radiograph is mostly a large-scale image larger than 1000000 pixels. The large-scale chest radiograph needs to be zoomed to (224 * 224) small-scale image to be applied to the diagnosis of lung diseases by vgg19, resnet50 and densenet169. However, this kind of operation will affect the pathological features in the chest radiograph. High-scale image scaling will lead to a large number of details loss (Figure\ref{fig4}), and at the same time, it will cause the learning and judgment ability of neural network to decline. So, if we increase the input size of the network, what will be the impact on the network structure itself? When the input size of the network is expanded, the input data in the feature level \cite{23} of the network will increase in addition to the number of operations, and have no other significant impact, but its output feature size will increase significantly. If the image size is 512 * 512, in VGg, the size of this feature will be increased from 512 * 7 * 7 to 512 * 16 * 16. At this point, if you use flat to flatten all vectors, and then use the full connection layer, you will have two problems: first, the increase of the number of vectors will make the classifier The parameters of level increase greatly, thus increasing the computational complexity, and a large number of parameters will accelerate the occurrence of over fitting; secondly, only increasing the size of the input image, but not changing the image network structure, will ultimately result in the perception field at the highest level of the model less than the image range, which has been proved to be inefficient in previous studies \cite{23}.

\begin{figure}[!t]
	\centerline{\includegraphics[width=\columnwidth]{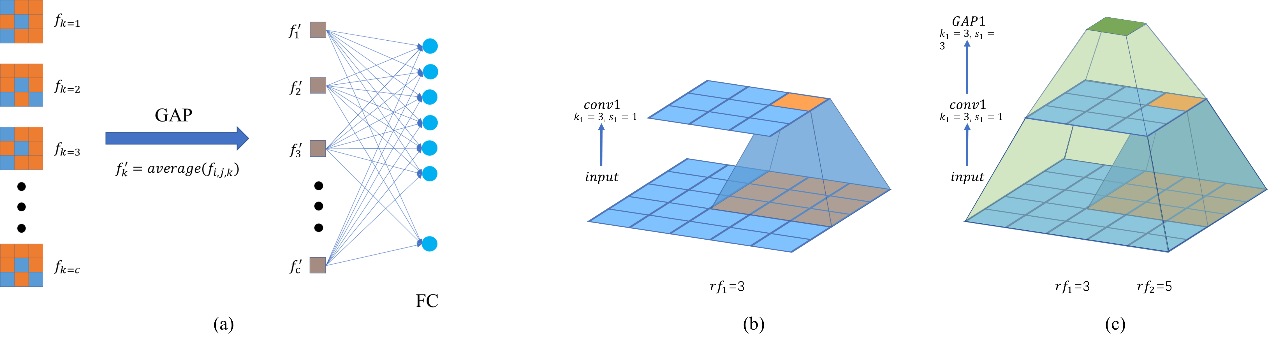}}
	\caption{(a) The schematic diagram of GAP, (b) the receptive field area before the gap structure is increased, (c) the receptive field area after the gap structure is increased.}
	\label{fig5}
\end{figure}

In order to solve the problem, we referred to the idea of global average pooling in \cite{24} and carried out global average pooling (GAP) on the last output eigenvector of our convolution layer. Let $f_{k}^{'}$ be the result of channel $k$ passing through GPA, and $i$ and $j$ be the coordinates of feature pixels in channel $k$. The GPA formula is expressed as: $$f_{k}^{'}=average(f_{i,j,k})$$
It can be concluded from the formula that after the features pass-through this structure, the size of each channel in it will become 1*1. So the input size of the classifier level will be greatly reduced and the number of parameters will remain stable (Figure\ref{fig5}(a)). Next, we will discuss the change of the receptive field. Let $rf_{n}$ be the receptive field of the $n-th$ layer of the network, $k_{n}$ be the kernel of the $n-th$ layer, and $s_{n}$ be the step size of the kernel. The calculation formula of the receptive field is expressed as: $$rf_{n}=rf_{n-1}*k_{n}-(k_{n}-1)*(rf_{n-1}-\prod_{i=1}^{n-1}s_{i})$$
As shown in Figure\ref{fig5} (b) and (c), after adding GAP to the convolution layer, the size of the newly formed receptive field will increase with the increase of the size of the input picture. This adaptive method can meet the high-efficiency requirements of the neural network model in \cite{23}.

\subsubsection{SE structure and attention mechanism}

\begin{figure}[!t]
	\centerline{\includegraphics[width=\columnwidth]{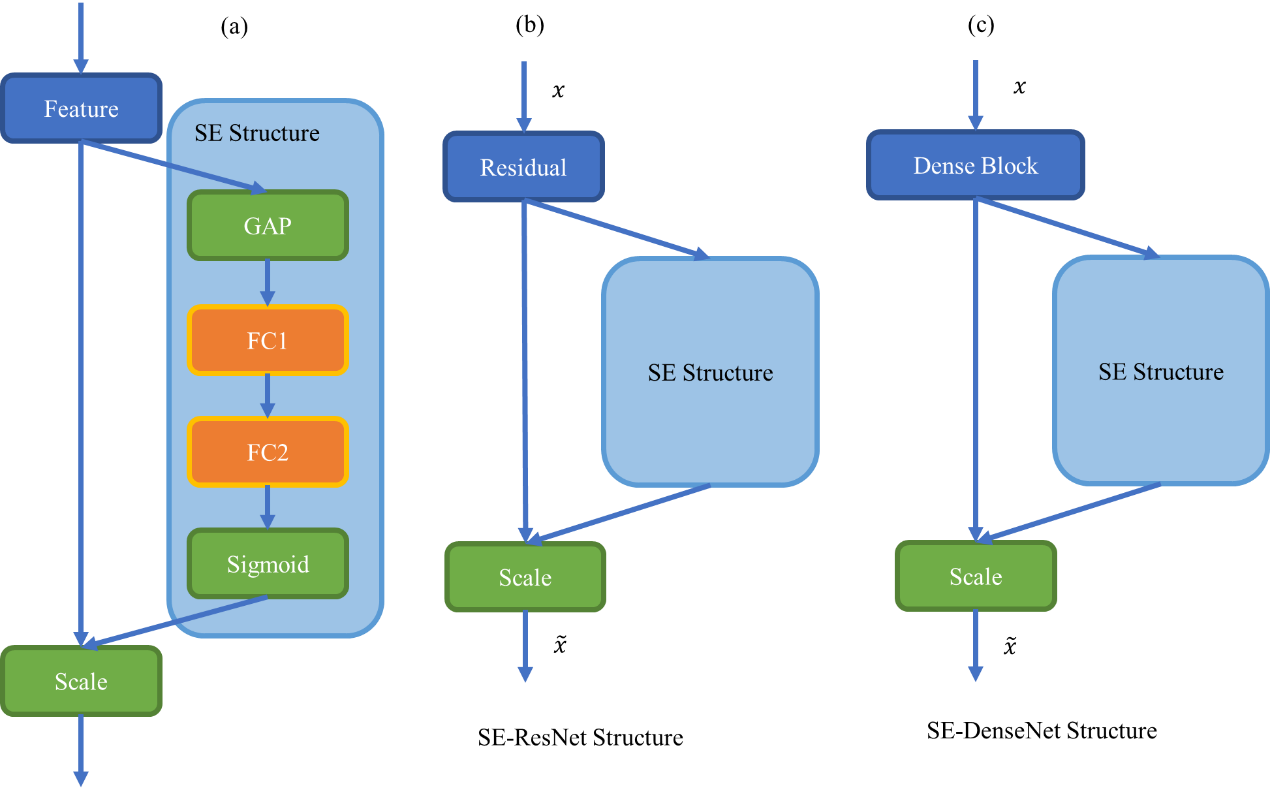}}
	\caption{(a) Squeeze-and-Exception structure, (b) SE-ResNet structure, (c) SE-DenseNet structure.}
	\label{fig6}
\end{figure}

Another improvement we have made to the network is to add attention to the structure of the network model. For chest image, even trained doctors need to analyze every detail of the image to determine the core of the problem, so as to get a more accurate judgment. But the neural networks also need such a mechanism to improve the accuracy of their judgment. Jie Hu et al. \cite{25} put forward a kind of network structure that has the function of attention and the name is Squeeze-Excitation which is hereinafter referred to as the SE module. SE module mainly includes Squeeze and Excitation two operations. It can apply to any mapping. In the convolution algorithm, a convolution kernel $V=[v_{1},v_{2},v_{3},...,v_{C}]$. $v_{C}$ means the $C-th$ convolution kernel. The output can be expressed as $U=[u_{1},u_{2},u_{3},...,u_{C}]$: $$u_{c}=v_{c}X=\sum_{s=1}^{C^{'}}v_{C}^{s}x^{s}$$
$v_{C}^{s}$ represents a three-dimensional convolution kernel. Because the convolution result of each channel is summed, the characteristic relation of each channel it learns is mixed with the spatial relation. The SE structure is to extract the mixture and make the model directly learn the feature relationship in the channel. First, the Squeeze structure encodes the entire spatial feature in a channel as a global feature and uses global average pooling to achieve this, which solves the problem that U is difficult to obtain enough information to extract the relationship between channels because convolution only operates in a local space. Then, in order to reduce model complexity and improve generalization ability, the squeeze was followed by two fully connected layers (FC). Between them, the first FC layer plays a role of dimensionality reduction, and the Dimension reduction coefficient is r which is a super parameter. Then, ReLU is used to activate the second FC layer to restore the original dimension. Excitation with Sigmoid Gating mechanism make the network learn the nonlinear relationship between each channel, $$s=F_{ex}(z,W)=\sigma(g(z,W))=\sigma(W_{2}ReLU(W_{1}z)), W_{1} \in R^{\frac{C}{r}*C}, W_{2} \in R^{C*\frac{C}{r}}$$
Finally, the activation value of each channel learned (sigmoid activation, value 0~1) is multiplied by the original feature on U to change the network's attention to each channel. This mechanism can be applied to ResNet and DenseNet with fusion layers, as shown in Figure\ref{fig6}. Considering that VGG has no merge layer. It is a network that uses the parameter layer to directly try to learn the mapping between input and output. If the attention structure is added, it will change the structure of the entire network model, which may also lead to its loss of stability. So we did not do this for VGG.

\subsection{Data promotion method}

\subsubsection{Adaptive histogram equalization with limited contrast}
Histogram equalization (HE) is often used to enhance the grayscale image in the medical field to obtain a clearer and more reliable image. Histogram equalization can improve the image contrast and make some previously imperceptible texture features clearly visible.

\begin{figure}[!t]
	\centerline{\includegraphics[width=\columnwidth]{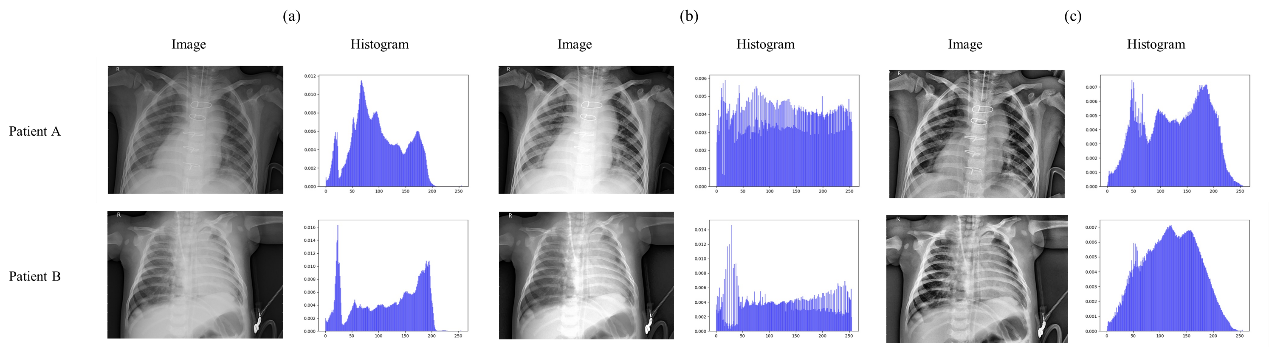}}
	\caption{Histogram equalization enhancement effect. (a) Original image and histogram, (b) histogram equalization(HE) effect and histogram, (c) The effect of histogram equalization with limited contrast and histogram.}
	\label{fig7}
\end{figure}

If a grayscale image has N pixels and the grayscale range is $[0, m-1]$, then the histogram equalization fraction of the image can be expressed as: $$s=T(r_{k})=\sum_{j=0}^{k}\frac{n_{j}}{N}=\sum_{j=0}^{k}P_{r}(r_{j})$$
$T(r_k )$  represents the mapping function of the original image and the equalization image at the $k-th$ grayscale. s is the cumulative distribution function of the gray level of the image. $n_{j}$ is the number of pixels at the gray level of $j$. $P_{r}(r_{j})$ represents the probability of the $j-th$ grayscale of the image. $\sum_{j=0}^{k}P_{r}(r_{j})$ represents the probability of grayscale from $0$ to $k$. The use of HE in the chest radiograph can enhance the overall contrast of the X-ray film to a certain extent, but the image after processing is excessively enhanced in the light and dark changes, resulting in the unnatural contrast as shown in Figure\ref{fig7}(b). It can be seen that the use of HE processing will often lead to the loss of target details, excessive background enhancement, and noise amplification. By limiting the maximum slope of cumulative distribution function (CDF), the limitation of the histogram equalization (HE) algorithm is overcome, and the random noise introduced in the process of histogram equalization is eliminated. The CLAHE algorithm differs from the ordinary adaptive histogram equalization in that it limits the contrast amplitude. Adaptive histogram equalization first divides the image into non-overlapping blocks and then use HE to each block. In CLAHE, the contrast amplitude must be limited for each small region. From HE, it can be seen that for any grayscale r of the image, the relation between the mapping curve T and CDF is: $$T(r)=\frac{K}{L}CDF(r)$$
The K is the highest grayscale value and L is the number of pixels. To limit the contrast amplitude, it actually also is to limit the cumulative histogram $CDF(r)$ of the slop. And the relationship between the $CDF(r)$ and the gray histogram $\int Hist(r)dr$ is: $$CDF(r)=\int Hist(r)dr$$
This formula indicates that limiting the slope of CDF is limiting the amplitude of the gray histogram. Using this technique in chest radiographs can improve the effectiveness of chest radiographs obviously. After using CLAHE, the boundary between bones and bones and the boundary between bones and organ tissues become clearer, and some detailed textures can be clearly seen, as shown in Figure\ref{fig7}(c)

\subsubsection{MoEx}

\begin{figure}[!t]
	\centerline{\includegraphics[width=\columnwidth]{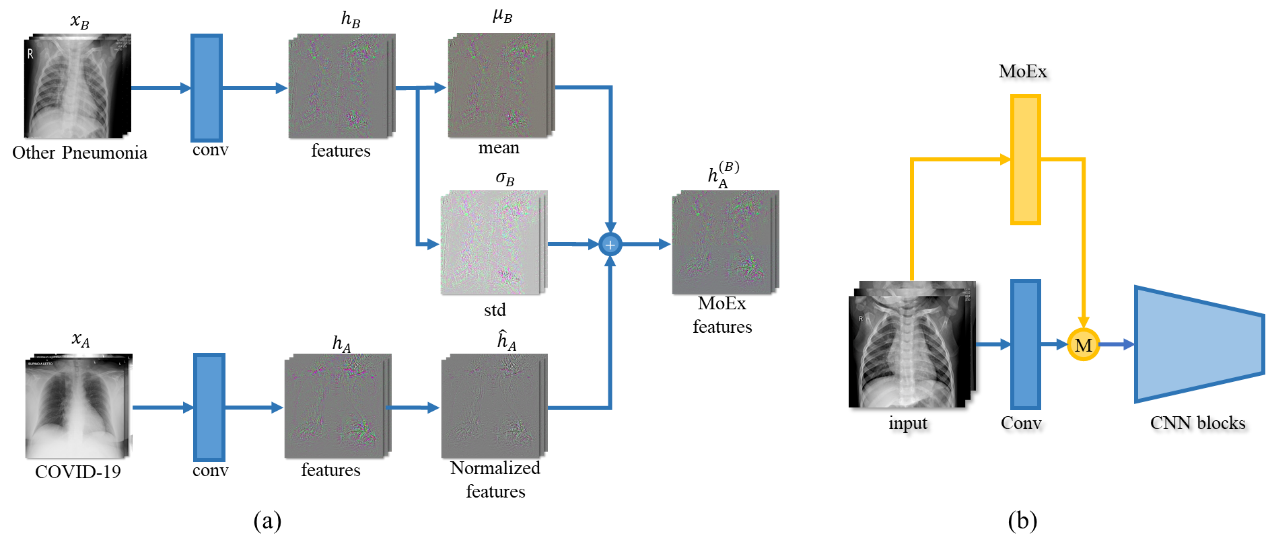}}
	\caption{After convolution, MoEx fused the features of the image, which enhanced the data from another dimension.}
	\label{fig8}
\end{figure}

MoEx \cite{22} is an algorithm to enhance image features in the network reasoning process. Unlike the traditional methods of data augmentation, MoEx mixes two different parts: the standardized features of one instance and the feature matrix of the other. In the feature space, this asymmetric combination enables the network to capture and smooth the different directions of the decision boundary, which is not covered by previous data augmentation methods. The function of normalization can be expressed as $F$, which takes the features $h_{i}^{l}$ of the ith input $x_{i}$ at layer $l$ and produces three outputs, the normalized features $\hat{h_{i}}$, the first-moment $\mu_{i}$, and the second-moment $\sigma_{i}$: $$(\hat{h_{i}^{l}},\mu_{i}^{l},\sigma_{i}^{l})=F(h_{i}^{l})$$
Since this method is applied in the same way at each layer, the '$l$' superscript is removed from the following formula to simplify the notation. Suppose the network inputs two different kinds of samples are $x_{A}$ and $x_{B}$, intra-instance normalization decomposes the features of input $x_{A}$ at layer $l$ into three parts, $\hat{h_{A}}$,$\mu_{A}$,$\sigma_{A}$ and $x_{B}$ into $\hat{h_{B}}$,$\mu_{B}$,$\sigma_{B}$, In order to encourage the network to utilize the moments, MoEx use the two images and combine them by injecting the moments of the image $x_{B}$ into the feature representation of the image $x_{A}$ $$h_{A}^{(B)}=F^{-1}(\hat{h_{A}},\mu_{B},\sigma_{B})$$
$h_{A}^{(B)}$ contain the moments of image B hidden inside the features of image A. In order to encourage the neural network to pay attention to the injected features of B MoEx modify the loss function to predict the class label $y_{A}$ and also $y_{B}$, up to some mixing constant $\lambda \in [0,1]$. The loss becomes a straight-forward combination $$\lambda*l(h_{A}^{(B)},y_{A})+(1-\lambda)*l(h_{A}^{(B)},y_{B})$$
MoEx enhanced the features of the image and improved the data from another dimension (Figure\ref{fig8}). This method can be superimposed with the CLAHE used in this paper. We applied such a data augmentation to ResNet50 and DenseNet169, and added MoEx structure after the first convolution layer of ResNet50 and DenseNet169 (Fig.8(b)). So we mixed two different types of pathological features, and modified the final loss function to adapt to such mixed features.

\subsubsection{Use U-Net to segment lung area}
U-Net is an improvement based on the VGG network. The images go through each convolution layer of the VGG to obtain the feature map. Then transposed convolution is used to the feature map and the pixel-level classification is carried out in the last layer. U-Net with full convolution network is adopted has achieved good results in medical image segmentation. X-ray films of lungs usually have the nonpathologic features, such as the letters used to mark the direction or other information marked with words. In the case of insufficient data, this often misleads network training to make the network extracts nonpathologic features mistakenly. We've shown that in our later assessment. In order to make the network ignore these nonpathologic features in the diagnosis process, we used U-Net \cite{9} to segment the training data in advance, and only retained the images of the double-lung part.

\subsection{Grad-CAM}
\begin{figure}[!t]
	\centerline{\includegraphics[width=\columnwidth]{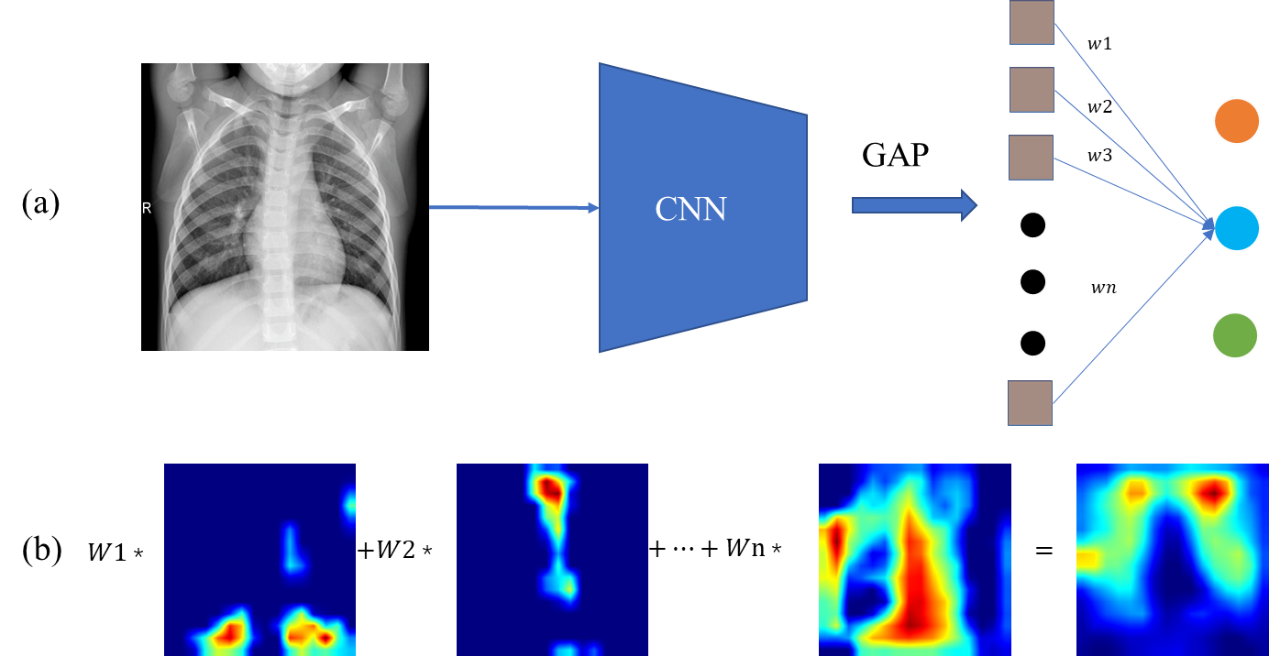}}
	\caption{Schematic diagram of Grad-CAM visual image.}
	\label{fig9}
\end{figure}

The study on the interpretation of the image feature regions learned by neural networks is one of the more difficult topics at present. In 2016, Bolei Zhou et al. proposed the visualization method of CAM \cite{26}. This method adopts the idea of NIN \cite{24} and uses the global average pooling layer to replace the fully connected layer. After passing through the convolutional neural network, the output of the last convolutional layer is globally averaged pooling to obtain a vector with the same length as the number of feature graphs. Between this vector and the correct class of the three classification results are the weight of $W1,W2,...,Wn$. These weights represent the weighting coefficients of different feature graphs. The heat map with the same size as the feature graphs is generated by adding these feature graphs according to the weighting coefficients. Then, the interpolation method is used to carry out upsampling to obtain a heat map of the same size as the original Figure\ref{fig9} is the schematic diagram of CAM. Grad-CAM is improved on the basis of CAM, and the weight of each feature graph is obtained by calculating the gradient information flowing into the last convolutional layer of CNN, which makes Grad-CAM applicable to any convolutional neural network. The use of visualization method in the chest radiographs can make the network provide the classification results and the classification basis at the same time so that the classification results are more reliable. To a certain extent, it can help doctors better understand the chest radiographs.

\section{Experiment and evaluation}
\begin{table}
	\label{tab3}
	\renewcommand\tabcolsep{8.0pt} % 调整表格列间的宽度
	\centering
	\caption{Dataset1 and dataset2 data enhancement methods and quantity}
\begin{tabular}{llllll}
	\hline
	\multicolumn{2}{c}{\textbf{Dataset I}} &  & \multicolumn{3}{c}{\textbf{Dataset II}} \\ \cline{1-2} \cline{4-6} 
	& \textbf{CLAHE} &  &  & \textbf{U-Net} & \textbf{CLAHE} \\ \hline
	\textbf{noraml} & 466 &  & COV19 & 105 & 105 \\
	\textbf{becteria} & 860 &  & others & 165 & 165 \\
	\textbf{virus} & 433 &  &  &  &  \\ \hline
\end{tabular}
	
\end{table}
The build and training of Cascade-SEMEnet is divided into two parts. In the first part, we evaluated the baseline performance of VGG19 \cite{27}, ResNet50 \cite{28}, and DenseNet169 \cite{29}, and compared the impact of different network structures and data processing methods on network performance. In the second part, due to the small amount of data, the network is prone to overfitting. Therefore, we carried out histogram equalization for all dataset2 and used MoEx to enhance image features in the experiment, which increased the diversity of images and reduced the occurrence of overfitting. In the second part, we also used U-Net to segment the lung region, excluding the influence of non-pathological features on network training. Since there is a big difference between the chest radiographs and the ImageNet, which was also discussed in 3-1 of this paper, the experiment did not use the pre-trained weights initialization network on the ImageNet, but adopted He-uniform: $W_{i,j}\sim U(-\sqrt{\frac{6}{n_{in}}},\sqrt{\frac{6}{n_{in}}})$ to initialize the convolution layers of the network model. In the evaluation phase, we compared the accuracy of the networks with different structures and enhancement methods on the test set and drew the ROC curve and confusion matrix to visualize the evaluation results. We used Grad-CAM to mark the pixels in the chest radiograph belonging to the classification basis and draw the thermal map to provide the corresponding diagnostic basis.
\subsection{Diagnosis of pneumonia infection type}
\begin{figure}[!t]
	\centerline{\includegraphics[width=\columnwidth]{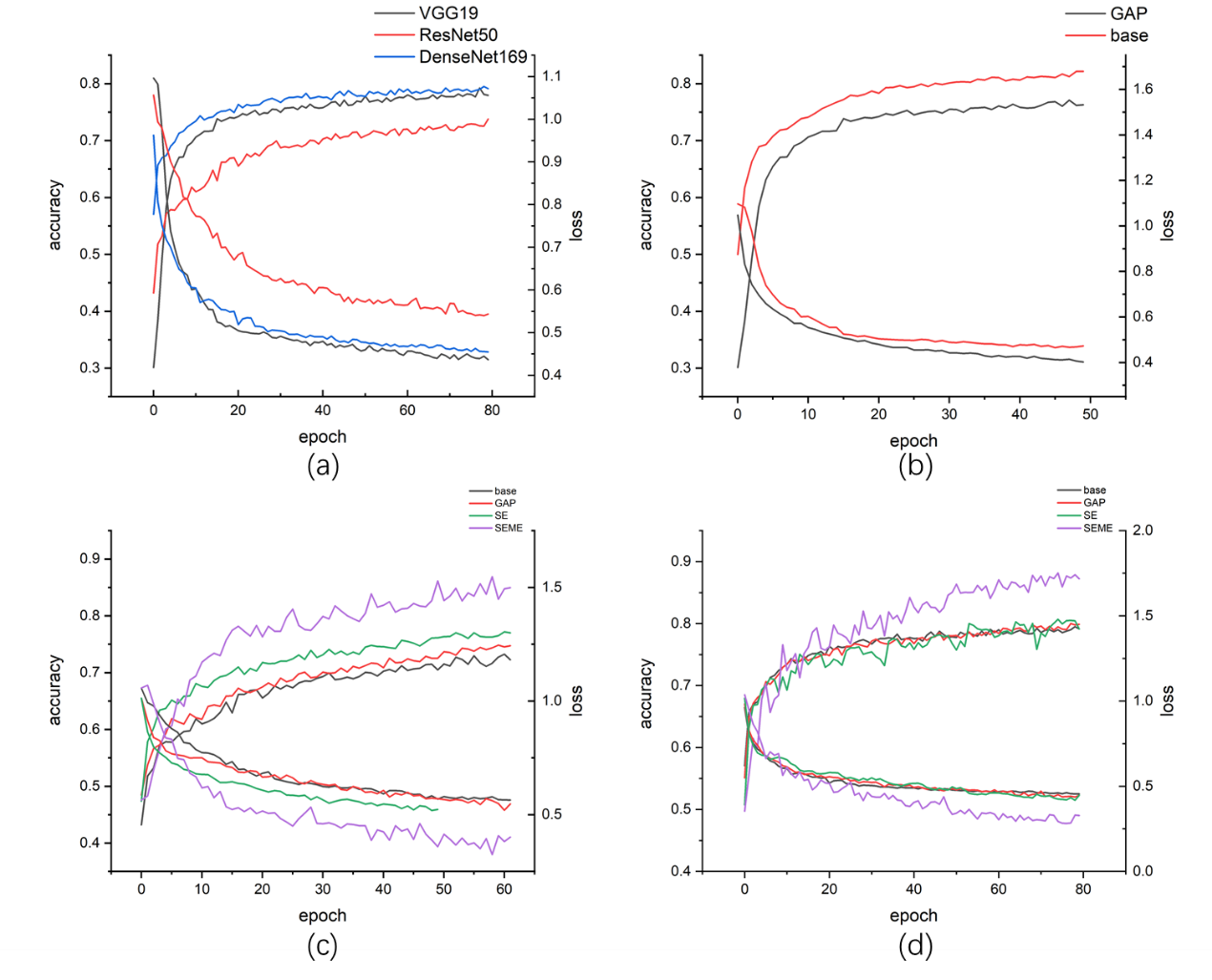}}
	\caption{(a) Training curve of VGG19, ResNet50, DenseNet169 on dataset1, (b) training curve of VGG19 and VGG19-GAP on dataset1, (c) training curve of ResNet50, ResNet50-GAP, SE-ResNet50, SEME-ResNet50 on dataset1, (d) training curve of DenseNet169, DenseNet169-GAP, SE-DenseNet169, SEME-DenseNet169 on dataset2.}
	\label{fig10}
\end{figure}

In the first stage, we used bacterial pneumonia, viral pneumonia, and normal chest radiographs to train the neural network model. The optimizers used in the training were SGD. The learning rate was set to 1e-4 and the momentum was set to 0.9. We first used unenhanced data to train three classic network structures to determine their baseline performance. The curve comparison during the training process is shown in Figure\ref{fig10}.
On the whole, the loss of the three kinds of curves is obviously convergent, which shows that the neural network can learn the characteristics from the chest radiograph to distinguish the types of pneumonia infection. Comparing the training curves of the three networks, it can be found that densenet169 with dense structure has the fastest convergence speed. After the first six rounds of training, it has 70\% accuracy, and the accuracy is still rising significantly. Vgg19 achieved 70\% accuracy in round 10, while resnet50 achieved 70\% accuracy in round 37 for the first time. After 80 rounds of training, the accuracy of vgg19, resnet50 and densenet169 on the test set is 69.7\%, 72.8\% and 77.5\% respectively (see Table\ref{tab4}), so it can be found that vgg19 has more serious over fitting. We added gap to the last convolution layer of vgg19, resnet50 and densenet169, adjusted the input size of the network to 512 * 512, and then retrained the training. The training curve is shown in Figure\ref{fig10} (b) (c) (d). After adding gap, the details of the input image become more abundant, the network also has a larger sense field, and the convergence speed of resnet50 and vgg19 has been significantly improved. The accuracy of vgg19 in the 10th round increased to 74\%, the accuracy in the 28th round reached 80\%, and the final accuracy in the test was 77.8\%. Compared with the initial network structure, the accuracy of vgg19 test set with gap increased by more than 8\%. Resnet50's accuracy increased to 72\% in round 37 and 74\% in the test set. However, after adding densenet169 into gap, the convergence rate has not been significantly improved. We speculate that densenet network's dense connection layer may dilute the network's attention to details and affect the convergence rate of the network. However, the accuracy of densenet with gap in the test set is significantly improved, reaching 80.9\%. Then, after adding Se structure to resnet50 and densenet169 networks, we train the network with the same data. In resnet50 with SE-structure, the speed of convergence has been significantly improved (Figure\ref{fig10} (c)), the accuracy of the 15th round of network training has reached 70\%, and the accuracy of test set has been increased to 81.6\%. In densenet169, the improvement of convergence speed is still not obvious. We think that the stability of convergence is also related to denseness of densenet structure itself. A large number of channel fusion slows down the speed of Se structure. But in the test set, the network structure has achieved better results, reaching 81.9\%.

In terms of data set enhancement strategy, we have processed histogram equalization for 40\% of training set data randomly, and Table\ref{tab3} has counted the number of enhanced data. At the same time, we add MOEX structure to se-resnet50 and se-densenet169, and use the algorithm of feature normalization to fuse the features of different kinds of lesions. After the image enhancement is added, the convergence speed of the two networks is improved more obviously. As shown in Figure\ref{fig10} (d), the accuracy of seme resnet50 in the 26th round is 81\%, and the final accuracy on the test set can reach 85.6\%, which is nearly 13\% higher than that of resnet50, and the accuracy of seme densenet169 in the 29th round for the first time is more than 80\%, and the accuracy in the test set is 80.3\%.
\begin{table}
	\label{tab4}
	\centering
	\caption{The accuracy and F1-score of the network on Test dataset1.}
	\renewcommand\tabcolsep{2.0pt} % 调整表格列间的宽度
	\begin{tabular}{cccccccc}
		\toprule[1pt]
		\multicolumn{4}{c}{\textbf{accuracy}} & \textbf{} & \multicolumn{3}{c}{\textbf{F1-score}} \\ \cline{1-4} \cline{6-8} 
		\textbf{} & \textbf{VGG19} & \textbf{ResNet50} & \textbf{DenseNet169c} & \textbf{} & \textbf{VGG19} & \textbf{ResNet50} & \textbf{DenseNet169} \\ \hline
		\textbf{base} & 69.69\% & 72.81\% & 77.50\% &  & 0.70 & 0.73 & 0.78 \\
		\textbf{GAP} & 77.81\% & 74.06\% & 80.94\% &  & 0.78 & 0.74 & 0.81 \\
		\textbf{SE} & - & 81.59\% & 81.87\% &  & - & 0.80 & 0.82 \\
		\textbf{SEME} & - & \textbf{85.62}\% & 80.31\% &  & - & \textbf{0.86} & 0.81 \\ \bottomrule[1pt]
	\end{tabular}
	
\end{table}
\begin{figure}[!t]
	\centerline{\includegraphics[width=\columnwidth]{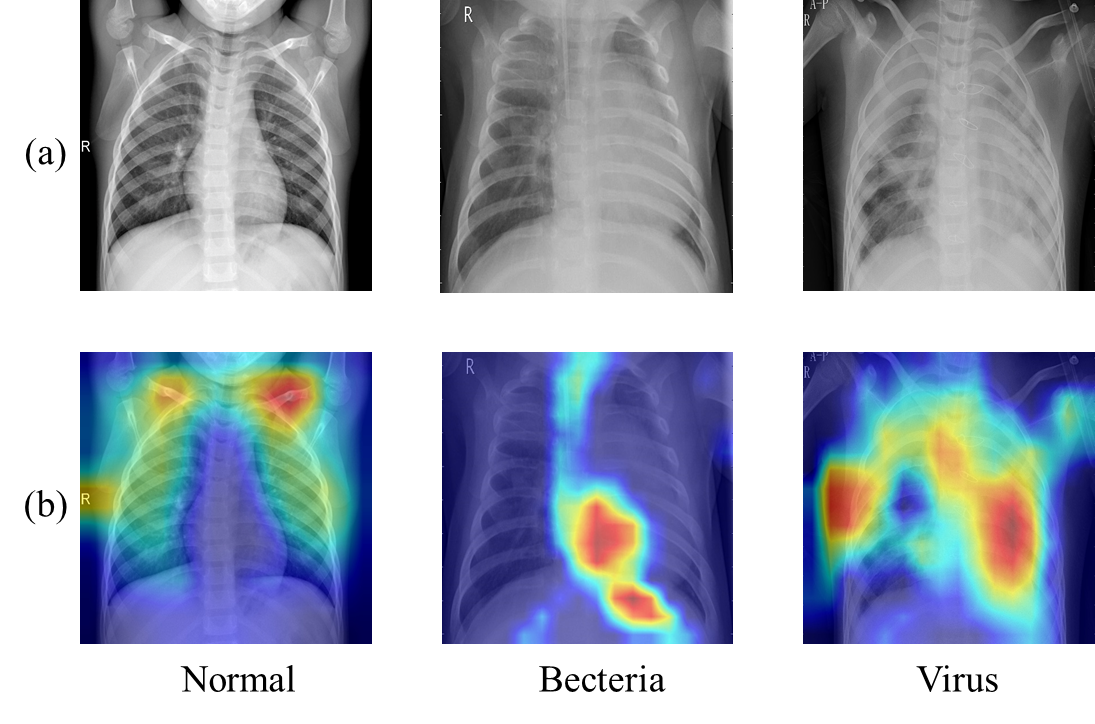}}
	\caption{The result of data visualization in dataset1 by Grad-CAM, (a) represents the original map, (b) overlays the original map on the Heatmap after Grad-CAM visualization.}
	\label{fig11}
\end{figure}
In order to verify the effectiveness of the network, we use the Grad-CAM method to visualize the judgment results of SEME-ResNet50 network, as shown in Figure\ref{fig11}, (a) is the original image put into the network, (b) is the chest thermal diagram visualized by the Grad-CAM method. From the visualization results, we can see that the focus areas judged by the network are basically concentrated in the position of the lung in the chest radiographs, in which the diagnosis basis of bacterial pneumonia is mainly located in the middle and lower part of the lung, and the basis of viral pneumonia is mainly concentrated in the central area of the lung.

\begin{figure}[!t]
	\centerline{\includegraphics[width=\columnwidth]{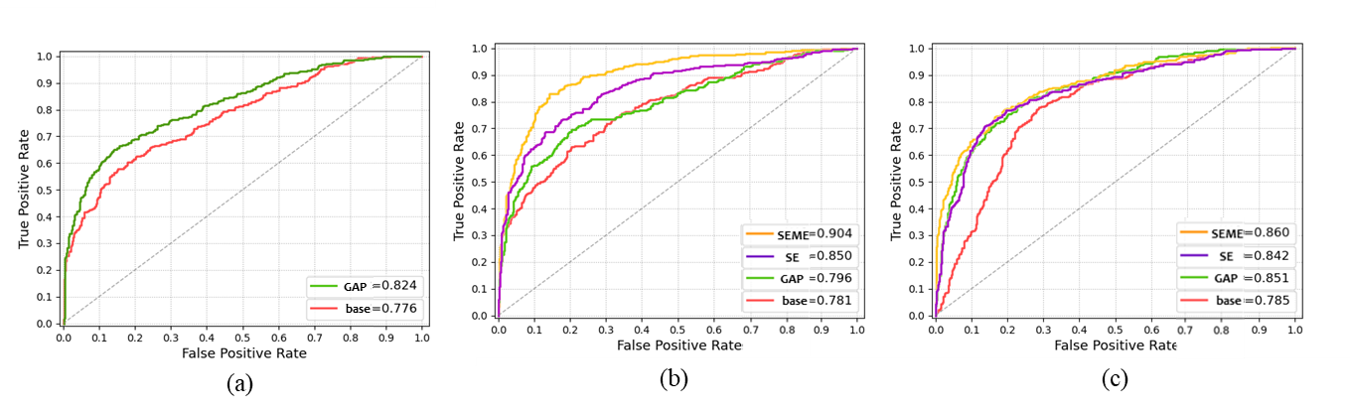}}
	\caption{(a) ROC curve of VGG19, VGG19-GAP on test dataset1, (b) ROC curve of ResNet50, ResNet50-GAP, SE-ResNet50, SEME-ResNet50 on test dataset1, (c) ROC curve of DenseNet169, DenseNet169-GAP, SE-DenseNet169, SEME-DenseNet169 on test dataset1.}
	\label{fig12}
\end{figure}
\begin{figure}[!t]
	\centerline{\includegraphics[width=\columnwidth]{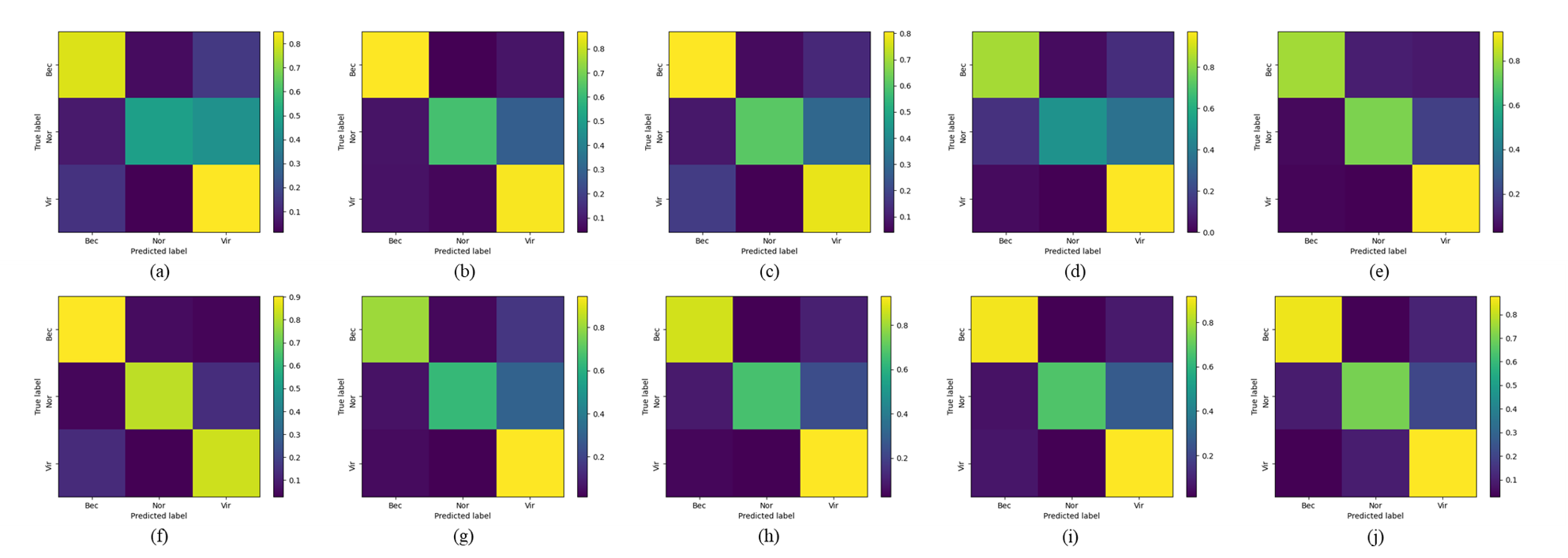}}
	\caption{(a) (b) confusion matrix of VGG19, VGG19-GAP on test dataset1, (c) (d) (E) (f) confusion matrix of ResNet50, ResNet50-GAP, SE-ResNet50, SEME-ResNet50 on test dataset1, (g) (H) (I) (J) confusion matrix of DenseNet169, DenseNet169-GAP, SE-DenseNet169, SEME-DenseNet169 on test dataset1.}
	\label{fig13}
\end{figure}

The evaluation results of each network on the test set are shown in Figure\ref{fig12} \ref{fig13}. Figure\ref{fig12}(a) is the ROC curve of VGG19 network after applying GAP and Figure\ref{fig13}(a) is its confusion matrix. It can be seen that after adding GAP into the structure, the AUC increases from 0.776 to 0.824, the confusion matrix is also obviously focused on the diagonal area, and the average f1-score increases from 0.70 to 0.78. In Figure\ref{fig12}(b) and Figure\ref{fig13}(b), the curve and confusion matrix after data promotion algorithm, GAP, and SE-Structure are added into ResNet50, while (c) is the ROC curve and confusion matrix of DenseNet169. It can be seen that after adding GAP and SE-Structure to the networks, the AUC of ResNet50 increased from 0.781 and 0.850 and of DenseNet169 increased from 0.785 and 0.851. The AUC of SEME-ResNet50 increased again to 0.904 and SEME-DenseNet169 increased to 0.860. The confusion matrix also converges to the diagonal area. The f1-scores of SEME-ResNet50 and SEME-DenseNet169 also increase greatly, from 0.73 to 0.86 and 0.78 to 0.81 respectively.

\subsection{Viral pneumonia and subdivision of COVID-19}
\begin{figure}[!t]
	\centerline{\includegraphics[width=\columnwidth]{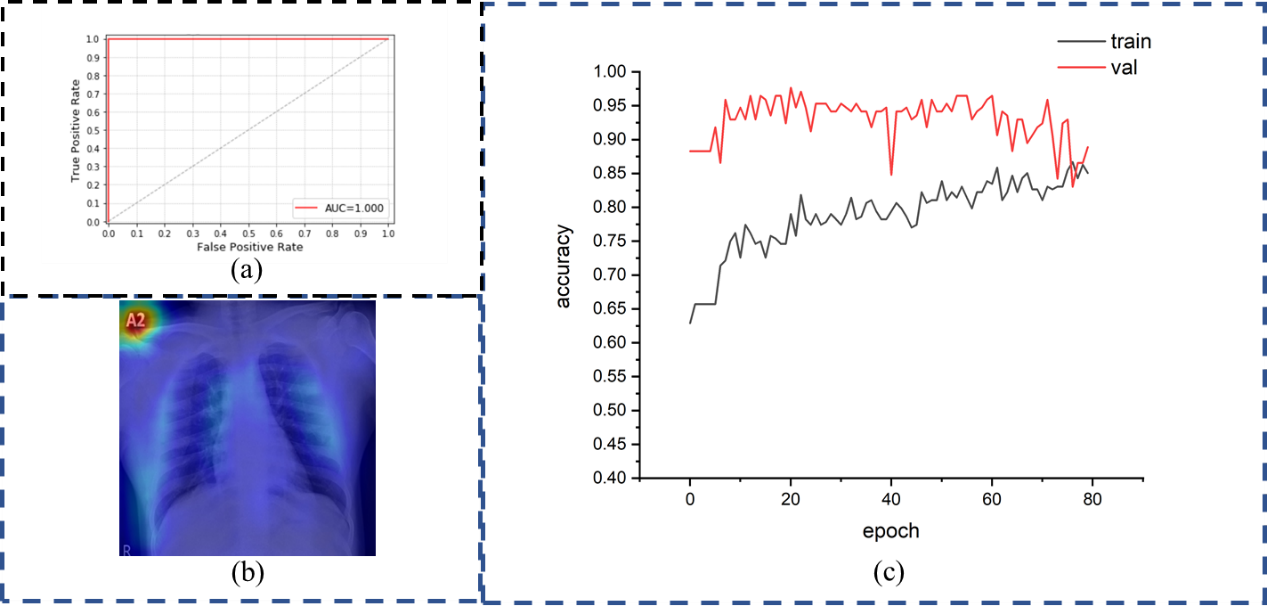}}
	\caption{(a) ROC curve on test dataset2, (b) visualization result of Grad-CAM on this network, (c) training curve of dataset2's original data on VGG19.}
	\label{fig14}
\end{figure}

The next step of Cascade-SEMEnet in identifying viral pneumonia is to perform fine-grained classification of viral pneumonia to diagnose COVID-19. Also, we first used the raw data on VGG19, ResNet50, DenseNet169 for training. Due to the small amount of data, we randomly rotated the images during the training, limiting the rotation amplitude to $\pm 30^{\circ}$, which increased the sample distribution to a certain extent. However, the results of the training were quite unexpected: the curve showed that the fitting speeds of the data on the three networks were very fast, the accuracy of the verification set reached 97\%, and the AUC of the test set reached 1.0, as shown in Figure\ref{fig14} (a) (c). It was almost impossible under normal conditions. We visualized the input data using Grad-CAM method and found that the convergence area of the network converges to the label of the chest radiographs (Figure\ref{fig14}(b)). We guess that it is because the number of chest radiographs of dataset2 is small, which leads to the fact that the labels such as letters and Numbers in the chest radiographs cause interference to the networks and enable the network to learn the non-pathological features.

\begin{figure}[!t]
	\centerline{\includegraphics[width=\columnwidth]{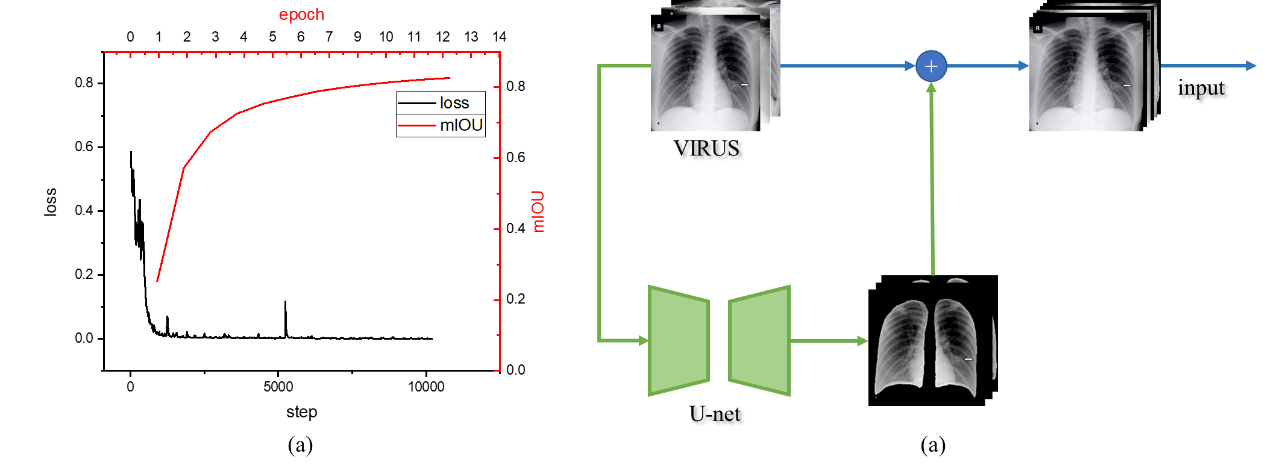}}
	\caption{(a) is the loss and mean IOU in the verification set during the U-Net training, (b) flow chart of U-Net processing and stacking data.}
	\label{fig15}
\end{figure}

To exclude other influences and enable the network to focus on the diagnosis of lung lesions, the U-Net was trained and used to segment the lung area of the chest radiographs in Dataset2. We used Adam optimizer in the training process of U-Net and we also use cosine annealing \cite{30} $ \eta _{t}=\eta_{min}^{i}+\frac{1}{2}(\eta_{max}^{i}-\eta_{min}^{i})(1+cos(\frac{T}{T_{i}})$. Loss in the training process and mean-iou in the verification set are shown in Figure\ref{fig15}(a). Figure\ref{fig15}(b) is a flow chart of U-Net processing and superimposing data.

Due to the small number of data, we used CLAHE to enhance all the data in dataset2, and then used U-Net to segment the original data and all the data after the promotion, and added the data before and after the segmentation as the training set. Meanwhile, we added MoEx structure to the network to further improve the data. In order to avoid network convergence point fall into a local optimal solution to accelerate network convergence, we took the training strategy of cosine annealing of learning rate and limited the maximum learning rate $\eta_{max}^{i}$ to 0.1 and the minimum learning rate $\eta_{min}^{i} to ie-8$.

\begin{figure}[!t]
	\centerline{\includegraphics[width=\columnwidth]{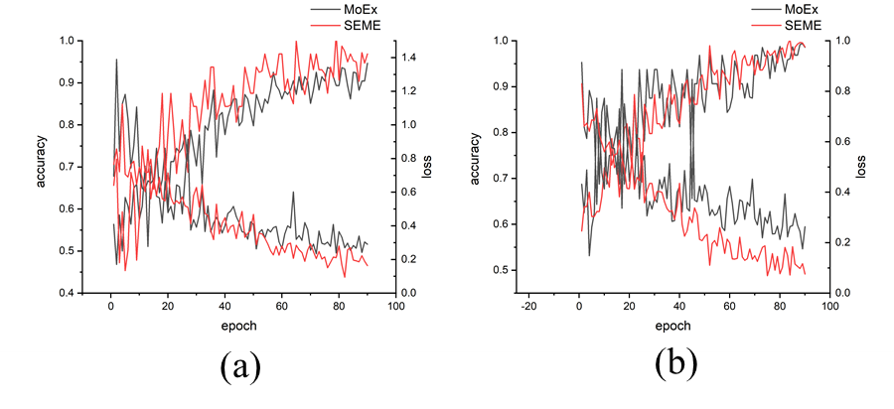}}
	\caption{(a) Training curve of MoEx-ResNet50 and SEME-ResNet50 in dataset2, (b) training curve of MoEx-DenseNet169 and SEME-DenseNet169 in dataset2.}
	\label{fig16}
\end{figure}

\begin{figure}[!t]
	\centerline{\includegraphics[width=\columnwidth]{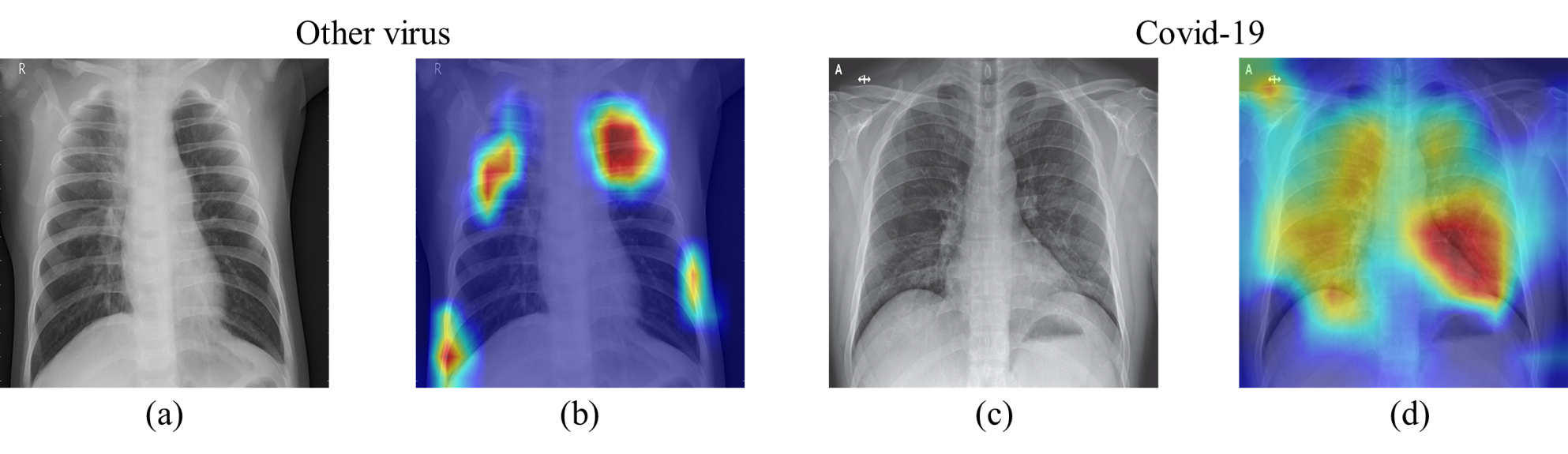}}
	\caption{The result of data visualization in dataset2 by Grad-CAM, (a)(c) is the original map, (b)(d) is the Heatmap of the original map after the Grad-CAM visualization.}
	\label{fig17}
\end{figure}

From the training curve of the network (Figure\ref{fig16} (a) (b)), it can be seen that the convergence speeds of SEME-ResNet50 and SEME-DenseNet169 are significantly faster than that of MoEx-ResNet50 and MoEx-DenseNet169. In the test(see Table\ref{tab5}), SEME-DenseNet169 achieves an accuracy rate of 97.1%.
\begin{table}
	\label{tab5}
	\centering
	\caption{The accuracy and F1-score of the network on Test dataset2.}
	\renewcommand\tabcolsep{2.0pt} % 调整表格列间的宽度
	\begin{tabular}{cccccc}
		\toprule[1pt]
		\multicolumn{3}{c}{\textbf{accuracy}} & \textbf{} & \multicolumn{2}{c}{\textbf{F1-score}} \\ \cline{1-3} \cline{5-6} 
		& \textbf{ResNet50} & \textbf{DenseNet169} & \textbf{} & \textbf{ResNet50} & \textbf{DenseNet169} \\
		\textbf{MoEx} & 94.28\% & 95.71\% &  & 0.94 & 0.96 \\
		\textbf{SE-MoEx} & 92.85\% & \textbf{97.14\%} &  & 0.93 & \textbf{0.97} \\ \bottomrule[1pt]
	\end{tabular}
	
\end{table}
The performance of the network on Grad-CAM is shown in Figure\ref{fig17}(b)(d). It can be found that when the network subdivides COVID-19, the discriminant criterion occupies almost the entire lung. The article\cite{11} analyzed chest radiographs of patients with COVID-19 and found that the most common chest radiographs were airspace opacities, whether described as consolidation or, less commonly, GGO. The networks we trained also noticed these characteristics. The visualization of COVID-19 in Figure\ref{fig17}(d) shows that the network focused on the lower right lung region in the process of judging the type of chest radiographs. In the original images (Figure\ref{fig17}(c)), the lesion described in the paper did appear in this region.

\begin{figure}[!t]
	\centerline{\includegraphics[width=\columnwidth]{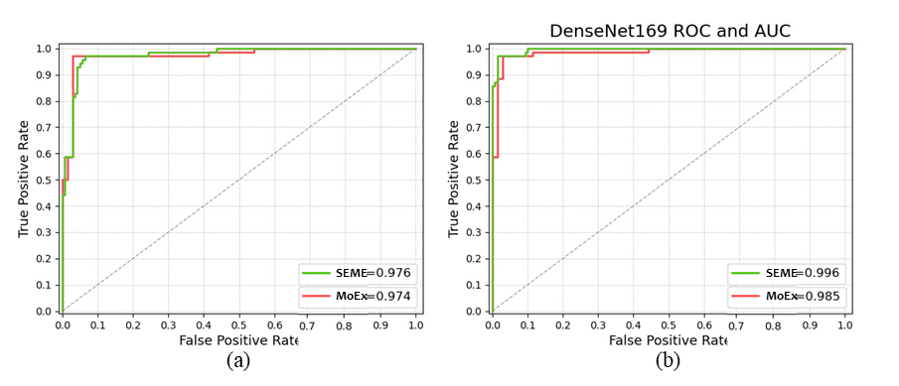}}
	\caption{(a) ROC curve of MoEx-ResNet50 and SEME-ResNet50 on test dataset2, (b) ROC curve of MoEx-DenseNet169 and SEME-DenseNet169 on test dataset2.}
	\label{fig18}
\end{figure}

\begin{figure}[!t]
\centerline{\includegraphics[width=\columnwidth]{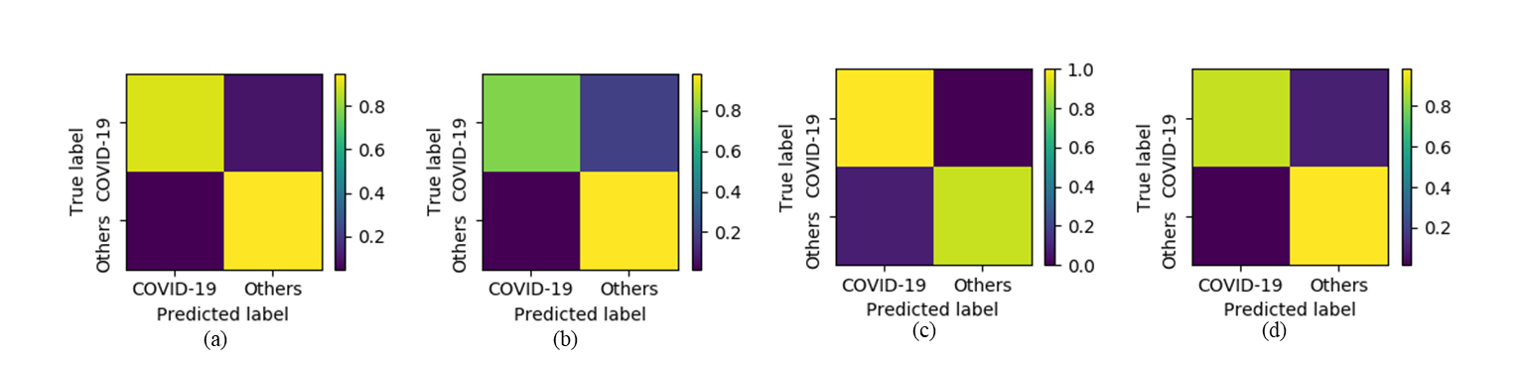}}
\caption{(a), (c) the confusion matrix of MoEx-ResNet50 and SEME-ResNet50 on test dataset2, (b), (d) the confusion matrix of MoEx-DenseNet169 and SEME-DenseNet169 on test dataset2.}
\label{fig19}
\end{figure}

Figure\ref{fig18}\ref{fig19} are the evaluation result of the network on the test set. Although there are few data and there may be some errors in the evaluation, but this result still has a certain reference value. It can be seen that the AUC of SEME-ResNet50 and SEME-DenseNet169 is improved to a certain extent compared with that of MoEx-Resnet50 and MoEx-DenseNet169. In the confusion matrix of the four networks (Figure\ref{fig19}), SEME-ResNet50 (e) and SEME-DenseNet169 (f) also converge to MoEx- ResNet50 (c) and MoEx- DenseNet169 (d). The maximum AUC of SEME-DenseNet169 can reach 0.996, and the F1-score of the test set reaches 0.97.

\section{Conclusion}
In this paper, the Cascade-SEMEnet composed of SEME-ResNet50, which detects the type of pulmonary infection, and DenseNet169, which is used for subdivision of viral pneumonia, was proposed to assist doctors in the diagnosis of pulmonary lesions and the recent outbreak of COVID-19 and a provide diagnostic basis for doctors. We used GAP to improve the network structure of ResNet and DenseNet, effectively making use of the pathological details of the images, increasing the receptive field of the network, adding SE-Structure to the network structure, and using the Attention mechanism for its characteristic channels. Experiments show that these structures can effectively improve the performance of the network. After adding GAP to VGG19, ResNet50, and DenseNet169, the evaluation of all networks on the dataset1 test set was significantly improved. After adding SE-Structure to ResNet50 and DenseNet169, although the convergence speed of SE-DenseNet169 was not significantly improved, the accuracy of SE-Resnet50 and SE-DenseNet169 on dataset1 test set was significantly improved, the largest increase was 9\%, compared with the original network. In order to make the network focus on the lung lesions in the training process and avoid the neural network from learning wrong information, we trained U-Net to segment the lung area of the chest radiographs, and put the segmented chest radiographs and the original chest radiographs into the network for training. CLAHE and MoEx methods also have good effects. The data enhanced by CLAHE are added into the training set, and the addition of MoEx structure in the network training can significantly increase the convergence speed of the network. In the evaluation of dataset1 test set, compared with the enhanced SE-ResNet50 network, the accuracy of SEME-ResNet50 is improved by more than 4\%. The accuracy of the dataset2 test set of SEME-DenseNet169 increases by 1.3\%. Its ROC curve and AUC also have different degrees of improvement.

We used Grad-CAM to visualize the basis of the network to judge the type of chest radiograph lesions. When the network is used to distinguish the chest radiograph of patients with no lesions, bacterial pneumonia, and viral pneumonia, the focus area is mainly on both sides of the lungs. But the location of different lesions is not the same. The chest radiograph of normal people is always located on the whole lung area on both sides. The main basis of chest radiographs of patients with bacterial pneumonia is the middle and lower region. This might be because the network used the pleural effusion \cite{31} and mediastinal lymph node enlargement \cite{32} as the judgment basis. In the chest radiograph of a patient with viral pneumonia, the network uses the central region of the entire lung as a basis for judgment. In the process of subdivision of viral pneumonia, in order to distinguish COVID-19, the network mainly judged other viruses based on the areas on the upper and middle sides of the double lung. While in the case of COVID-19, the network focused on the judgment of frosted glass shadow and consolidation of the lung on the basis of almost all the areas of the double lung. But we found that even U-Net was applied to remove the annotation interference of dataset2, due to too little data, the label of the original image will also cause certain influence to the network by overlying the cut image with the original image to train. As a result, the judgment basis of the network for the fine-grained classification of COVID-19 contains a small part of the annotation.

In the future work, we will use larger data sets to train the network. At the same time, we will try to use the knowledge distillation to soften the label, optimize the performance of the model, so that the model can be better applied in the clinical auxiliary diagnosis. And we also can use this method for attention transfer and to detect the complications of pneumonia.

%\bibliographystyle{model2-names}
%\bibliography{refs}

%----------------------------------------------------------------------------------------
%	BIBLIOGRAPHY
%----------------------------------------------------------------------------------------

\renewcommand{\refname}{\spacedlowsmallcaps{References}} % For modifying the bibliography heading

\bibliographystyle{unsrt}

\bibliography{refs}

%----------------------------------------------------------------------------------------

\end{document}

%% file: structure.tex
\usepackage[
nochapters, % Turn off chapters since this is an article        
beramono, % Use the Bera Mono font for monospaced text (\texttt)
eulermath,% Use the Euler font for mathematics
pdfspacing, % Makes use of pdftex’ letter spacing capabilities via the microtype package
dottedtoc % Dotted lines leading to the page numbers in the table of contents
]{classicthesis} % The layout is based on the Classic Thesis style

\usepackage{arsclassica} % Modifies the Classic Thesis package

\usepackage[T1]{fontenc} % Use 8-bit encoding that has 256 glyphs

\usepackage[utf8]{inputenc} % Required for including letters with accents

\usepackage{graphicx} % Required for including images
\graphicspath{{Figures/}} % Set the default folder for images

\usepackage{enumitem} % Required for manipulating the whitespace between and within lists

\usepackage{lipsum} % Used for inserting dummy 'Lorem ipsum' text into the template

\usepackage{subfig} % Required for creating figures with multiple parts (subfigures)

\usepackage{amsmath,amssymb,amsthm} % For including math equations, theorems, symbols, etc

\usepackage{varioref} % More descriptive referencing

\theoremstyle{definition} % Define theorem styles here based on the definition style (used for definitions and examples)

\theoremstyle{plain} % Define theorem styles here based on the plain style (used for theorems, lemmas, propositions)

\theoremstyle{remark} % Define theorem styles here based on the remark style (used for remarks and notes)

\hypersetup{
colorlinks=true, breaklinks=true, bookmarks=true,bookmarksnumbered,
urlcolor=webbrown, linkcolor=RoyalBlue, citecolor=webgreen, % Link colors
pdftitle={}, % PDF title
pdfauthor={\textcopyright}, % PDF Author
pdfsubject={}, % PDF Subject
pdfkeywords={}, % PDF Keywords
pdfcreator={pdfLaTeX}, % PDF Creator
pdfproducer={LaTeX with hyperref and ClassicThesis} % PDF producer
}